\documentclass[useAMS,usenatbib,fleqn,usegraphicx]{mn2e}

\usepackage{mathrsfs}
\usepackage{amsmath}
\usepackage{amssymb}
\usepackage{amsfonts}
 
\newcommand{\de}{\mathrm d}
\newcommand{\om}{\Omega_m}
\newcommand{\ob}{\Omega_b}
\newcommand{\odm}{\Omega_\mathrm{DM}}

\newcommand{\ol}{\Omega_\Lambda}
\newcommand{\ho}{{H_0}}
\newcommand{\cinf}{{c_\infty}}
\newcommand{\lcdm}{$\Lambda$CDM}

\title[Measuring UDM with 3D Cosmic Shear]{Measuring Unified Dark Matter with 3D Cosmic Shear}
\author[S. Camera et al.]{Stefano Camera,$^{1,2,3}$\thanks{E-mail: camera@ph.unito.it (SC); tdk@roe.ac.uk (TDK); afh@roe.ac.uk (AFH); bertacca@pd.infn.it (DB); diaferio@ph.unito.it (AD)} Thomas D. Kitching,$^4$\footnotemark[1] Alan F. Heavens,$^4$\footnotemark[1]\newauthor Daniele Bertacca,$^{5,6,7}$\footnotemark[1] and  Antonaldo Diaferio$^{1,3,8}$\footnotemark[1]\\$^1$Dipartimento di Fisica Generale ``Amedeo Avogadro'', Universit\`a degli Studi di Torino, Torino, Italy\\$^2$Dipartimento di Fisica Teorica, Universit\`a degli Studi di Torino, Torino, Italy\\$^3$INFN\thanks{Istituto Nazionale di Fisica Nucleare}, Sezione di Torino, Torino, Italy\\$^4$SUPA\thanks{Scottish Universities Physics Alliance}, Institute for Astronomy, University of Edinburgh, Royal Observatory, Edinburgh, UK\\$^5$Dipartimento di Fisica ``Galileo Galilei'', Universit\`a di Padova, Padova, Italy\\$^6$INFN\footnotemark[2], Sezione di Padova, Padova, Italy\\$^7$Institute of Cosmology and Gravitation, University of Portsmouth, Portsmouth, UK\\$^8$Harvard-Smithsonian Center for Astrophysics, Cambridge MA, USA}

\begin{document}

\date{Accepted 00 -- 0000. Received 00 -- 0000; in original form \today}

\pagerange{\pageref{firstpage}-\pageref{lastpage}} \pubyear{2010}

\maketitle

\label{firstpage}

\begin{abstract}
We present parameter estimation forecasts for future 3D cosmic shear surveys for a class of Unified Dark Matter (UDM) models, where a single scalar field mimics both Dark Matter (DM) and Dark Energy (DE). These models have the advantage that they can describe the dynamics of the Universe with a single matter component providing an explanation for structure formation and cosmic acceleration. A crucial feature of the class of UDM models we use in this work is characterised by a parameter, $\cinf$ (in units of the speed of light $c=1$), that is the value of the sound speed at late times, and on which structure formation depends. We demonstrate that the properties of the DM-like behaviour of the scalar field can be estimated with very high precision with large-scale, fully 3D weak lensing surveys. We found that 3D weak lensing significantly constrains $\cinf$, and we find minimal errors $\Delta \cinf=3.0\cdot10^{-5}$, for the fiducial value $\cinf=1.0\cdot10^{-3}$, and $\Delta \cinf=2.6\cdot10^{-5}$, for $\cinf=1.2\cdot10^{-2}$. Moreover, we compute the Bayesian evidence for UDM models over the $\Lambda$CDM model as a function of $\cinf$. For this purpose, we can consider the $\Lambda$CDM model as a UDM model with $\cinf=0$. We find that the expected evidence clearly shows that the survey data would unquestionably favour UDM models over the $\Lambda$CDM model, for the values $\cinf\gtrsim10^{-3}$.
\end{abstract}


\section{Introduction}
Provided that General Relativity itself is accurate, a number of cosmological observations, e.g. the dynamics of galaxies and galaxy clusters and the Large-Scale Structure (LSS), provide us with ample evidence that most of the matter in the Universe is not made up of familiar baryonic matter, but of Dark Matter (DM) \citep{Zwicky:1933gu,Zwicky:1937zza,Dodelson:2001ux,Hawkins:2002sg,Spergel:2006hy,Riess:2006fw}. Moreover, the largest fraction of the energy budget in the present Universe is occupied by Dark Energy (DE), which seems to be accelerating the expansion of the Universe \citep{Riess:1998cb,Knop:2003iy,Riess:2004n,Riess:2006fw,Komatsu:2008hk}. The current concordance model, with radiation, baryons, cold DM and DE (in the form of a cosmological constant $\Lambda$) is known as the $\Lambda$CDM model.

It has been demonstrated \citep{Albrecht:2006um,2006ewg3.rept.....P} that the nature of the dark components of the Universe can be constrained to a high degree of accuracy by using wide and deep imaging surveys; weak lensing, in which the shear and redshift information of every galaxy is used, has the potential to constrain the equation of state of such dark components by using surveys such as Euclid\footnote{http://sci.esa.int/science-e/www/area/index.cfm?fareaid=102} \citep{2008SPIE.7010E..38R,Refregier:2010ss} or Pan-STARRS\footnote{http://pan-starrs.ifa.hawaii.edu}  \citep{2002SPIE.4836..154K,2002AAS...20112207K}. As a direct probe of the mass distribution, gravitational lensing is an excellent tool for cosmological parameter estimation, complementing Cosmic Microwave Background (CMB) studies. One of the most useful manifestations of gravitational lensing by intervening matter is the alignment of nearby images on the sky. Detection of DM on large scales through such cosmic shear measurements -- the small, coherent distortion of distant galaxy images due to the large-scale distribution of matter in the cosmos -- has recently been shown to be feasible.

At a statistical level, it has been shown \citep{Hu:1998az,Hu:1999ek} that there is some extra information on cosmological parameters which can be gained by dividing the sample into several redshift bins; this technique is known as weak lensing tomography. However, a more comprehensive representations of the shear field can be called 3D weak lensing \citep{Heavens:2003jx,Castro:2005bg,Heavens:2006uk,Kitching:2006mq}, in which, by using the formalism of spin-weighted spherical harmonics and spherical Bessel functions, one can relate the two-point statistics of the harmonic expansion coefficients of the weak lensing shear and convergence to the power spectrum of the matter density perturbations. Such a tool is relevant in view of the present and next generations of large-scale weak lensing surveys, which will provide distance information of the sources through photometric redshifts.

Recently, rather than considering DM and DE as two distinct components, it has been suggested the alternative hypothesis that DM and DE are two states of the same fluid. This has been variously referred to as ``Unified Dark Matter'' or ``quartessence'' models. Compared with the standard DM plus DE models (e.g. $\Lambda$CDM), these models have the advantage that we can describe the dynamics of the Universe with a single scalar field which triggers both the accelerated expansion at late times and the LSS formation at earlier times. Specifically, for these models, we can use Lagrangians with a non-canonical kinetic term, namely a term which is an arbitrary function of the square of the time derivative of the scalar field, in the homogeneous and isotropic background.

Originally this method was proposed to have inflation driven by kinetic energy, called $k$-inflation \citep{ArmendarizPicon:1999rj,Garriga:1999vw}, to explain early Universe's inflation at high energies. Then this scenario was applied to DE \citep{Chiba:1999ka,dePutter:2007ny,Linder:2008ya}. In particular, the analysis was extended to a more general Lagrangian \citep{ArmendarizPicon:2000dh,ArmendarizPicon:2000ah} and this scenario was called $k$-essence \citep[see also][]{Chiba:1999ka,Rendall:2005fv,Li:2006bx,Calcagni:2006ge,Babichev:2006cy,Fang:2006yh,Bazeia:2007df,Kang:2007vs,Babichev:2007dw,Babichev:2007tn,Ahn:2009xd}.

For zmodels, several adiabatic or, equivalently, purely kinetic models have been investigated in the literature: the generalised Chaplygin gas \citep{Kamenshchik:2001cp,Bilic:2001cg,Bento:2002ps,Carturan:2002si,Sandvik:2002jz}, the single dark perfect fluid with a simple two-parameter barotropic equation of state \citep{Balbi:2007mz,Quercellini:2007ht,Pietrobon:2008js} and the purely kinetic models studied by \citet{Scherrer:2004au}, \citet{Bertacca:2007ux}, \citet{Chimento:2009nj}. Alternative approaches have been proposed in models with canonical Lagrangians with a complex scalar field \citep{Arbey:2006it}.

One of the main issues of these UDM models is to see whether the single dark fluid is able to cluster and produce the cosmic structures we observe in the Universe today. In fact, a general feature of UDM models is the possible appearance of an effective sound speed, which may become significantly different from zero during the evolution of the Universe. In general, this corresponds to the appearance of a Jeans length (or sound horizon) below which the dark fluid does not cluster. Thus, the viability of UDM models strictly depends on the value of this effective sound speed \citep{Hu:1998kj,Garriga:1999vw,Mukhanov:2005sc}, which has to be small enough to allow structure formation \citep{Sandvik:2002jz,Giannakis:2005kr,Bertacca:2007cv} and to reproduce the observed pattern of the CMB temperature anisotropies \citep{Carturan:2002si,Bertacca:2007cv}.

In general, in order for UDM models to have a very small speed of sound and a background evolution that fits the observations, a severe fine tuning of their parameters is necessary. In order to avoid this fine tuning, alternative models with similar goals have been analysed in the literature: \citet{Piattella:2009kt} studied in detail the functional form of Jeans scale in adiabatic UDM perturbations and introduced a class of models with a fast transition between an early Einstein-de Sitter cold DM-like era and a later $\Lambda$CDM-like phase. If the transition is fast enough, these models may exhibit satisfactory structure formation and CMB fluctuations, thus presenting a small Jeans length even in the case of a non-negligible sound speed; \citet{Gao:2009me} explore unification of DM and DE in a theory containing a scalar field of non-Lagrangian type, obtained by direct insertion of a kinetic term into the energy-momentum tensor.

Here, we choose to investigate the class of UDM models studied in \citet{Bertacca:2008uf}, who designed a reconstruction technique of the Lagrangian, which allows one to find models where the effective speed of sound is small enough, and the $k$-essence scalar field can cluster (see also \citealt{Camera:2009uz}, \citealt{Camera:2010}). In particular, the authors require that the Lagrangian of the scalar field is constant along classical trajectories on cosmological scales, in order to obtain a background identical to the background of the $\Lambda$CDM model.

Here, we wish to investigate whether this class of UDM models can be scrutinised in realistic scenarios. Specifically, we compute the weak lensing signals expected in these models as they would be measured by a Euclid-like survey.

The structure of this paper is as follows. In Section~\ref{udm} we describe the UDM model we use in this work. In Section~\ref{3dlensing} we detail the theory of weak gravitational lensing on the celestial sphere, with a particular interest in the cosmic shear observable (Section~\ref{3dshear}). In Section~\ref{fisher} we outline the Fisher matrix formalism we use to calculate the expected statistical errors on cosmological parameters, and with the same formalism we compute the expected Bayesian evidence for UDM models over the standard $\Lambda$CDM model as a function of the sound speed parameter $\cinf$ (Section~\ref{B-evidence}). In Section~\ref{results} we present our results, such as the matter power spectrum obtained in these UDM models (Section~\ref{matterpowerspectrum}) and the corresponding 3D shear signal (Section~\ref{signal}); the parameter estimations for a Euclid-like survey are presented in Section~\ref{estimation}, while in Section~\ref{selection} we use the Bayesian approach to ask the data whether our UDM model is favoured over the $\Lambda$CDM model or not. Finally, in Section~\ref{conclusions}, conclusions are drawn.

\section{Unified Dark Matter models}\label{udm}
We consider a UDM model where the Universe is filled with a perfect fluid of radiation, baryons and a scalar field $\varphi(t)$, the latter mimicking both DM and DE in form of a cosmological constant. In particular, \citet{Bertacca:2008uf}, by using  scalar-field Lagrangians $\mathscr L(X,\varphi)$ with a non-canonical kinetic term, where\footnote{We use units such that $c=1$ and signature $\{-,+,+,+\}$, where Greek indices run over spacetime dimensions, whereas Latin indeces label spatial coordinates.}
\begin{equation}
X=-\frac{1}{2}\nabla^\mu\varphi\nabla_\mu\varphi,
\end{equation}
have outlined a technique to reconstruct UDM models such that the effective speed of sound is small enough to allow the clustering of the scalar field. Specifically, once the initial value of the scalar field is fixed, the scalar field Lagrangian is constant along the classical trajectories, namely $\mathscr L_\varphi=-\Lambda/(8\pi G)$, and the background is identical to the background of $\Lambda$CDM. In other words, the energy density of the UDM scalar field presents two terms
\begin{equation}
\rho_\mathrm{UDM}(t)=\rho_\mathrm{DM}(t)+\rho_\Lambda,
\end{equation}
where $\rho_\mathrm{DM}$ behaves like a DM component ($\rho_\mathrm{DM}\propto a^{-3}$) and $\rho_\Lambda$ like a cosmological constant component ($\rho_\Lambda=\mathrm{const.}$). Consequently, $\odm=\rho_\mathrm{DM}(a=1)/\rho_c$ and $\ol=\rho_\Lambda/\rho_c$ are the density parameters of DM and DE today, where $\rho_c$ is the present day critical density; hence, the Hubble parameter in these UDM models is the same as in $\Lambda$CDM,
\begin{equation}
H(z)=\ho\sqrt{\om{(1+z)}^3+\ol},
\end{equation}
with $\ho=100\,h\,\mathrm{km\,s^{-1}\,Mpc^{-1}}$ and $\om=\odm+\ob$, where $\ob=\rho_b/\rho_c$ is the baryon density in units of the critical density.

Now we introduce small inhomogeneities of the scalar field $\delta\varphi(t,\mathbf x)$, and in the linear theory of cosmological perturbations and in the Newtonian gauge, the line element is
\begin{equation}
\de s^2=-(1+2\Phi)\de t^2+a^2(t)(1+2\Psi)\de\mathbf x^2,\label{flrw}
\end{equation}
in the case of a spatially flat Universe, as supported by CMB measurements \citep[e.g.][]{Spergel:2006hy}. This scalar field presents no anisotropic stress, thus $\Phi=-\Psi$. With this metric, when the energy density of radiation becomes negligible, and disregarding also the small baryonic component, the evolution of the Fourier modes of the Newtonian potential $\Phi_\mathbf{k}(a)$ are described by \citep{Garriga:1999vw,Mukhanov:2005sc}
\begin{equation}
{v_\mathbf{k}}''+{c_s}^2k^2v_\mathbf{k}-\frac{\theta''}{\theta}v_\mathbf{k}=0,\label{eq-Mukhanov:2005sc-lcdm}
\end{equation}
where a prime denotes a derivative with respect to the conformal time $\de\tau=\de t/a$, $k=|\mathbf k|$ and
\begin{align}
v&\equiv\frac{\Phi}{\sqrt{\rho_\mathrm{UDM}+p_\mathrm{UDM}}}\label{udiphi-lcdm},\\\theta&\equiv\frac{1}{a\sqrt{1+\frac{p_\mathrm{UDM}}{\rho_\mathrm{UDM}}}};
\end{align}
here,
\begin{equation}
{c_s}^2(a)=\frac{{p_\mathrm{UDM}}_{,X}}{{\rho_\mathrm{UDM}}_{,X}}\label{c_s}
\end{equation}
is the effective speed of sound, where $_{,X}$ denotes a derivative w.r.t. $X$.

By following the technique outlined by \citet{Bertacca:2008uf}, it is possible to construct a UDM model in which the sound speed is small enough to allow the formation of the LSS we see today and is capable of reproducing the observed pattern of the temperature anisotropies in the CMB radiation. We choose a Lagrangian of the form
\begin{equation}
\mathscr L_\varphi\equiv p_\mathrm{UDM}(\varphi,X)=f(\varphi)g(X)-V(\varphi)\label{L_phi},
\end{equation}
with a Born-Infeld type kinetic term $g(X)=-\sqrt{1-2XM^{-4}}$ \citep{Born:1934gh}, where $M$ is a suitable mass scale. Such a kinetic term can be thought as a field theory generalisation of the Lagrangian of a relativistic particle \citep{Padmanabhan:2002sh,Abramo:2003cp,Abramo:2004ji}. It was also proposed in connection with string theory, since it seems to represent a low-energy effective theory of $D$-branes and open strings, and has been conjectured to play a role in cosmology \citep{Sen:2002nu,Sen:2002in,Sen:2002vv,Padmanabhan:2002sh}. By using the equation of motion of the scalar field $\varphi(t, \mathbf x)$ and by imposing that the scalar field Lagrangian is constant along the classical trajectories, i.e. $p_\mathrm{UDM}=-\rho_\Lambda$, we obtain the following expressions for the potentials
\begin{align}
f(\varphi)&=\frac{\Lambda \cinf}{1-{\cinf}^2}\frac{\cosh(\xi\varphi)}{\sinh(\xi\varphi)\left[1+\left(1-{\cinf}^2\right)\sinh^2(\xi\varphi)\right]},\\V(\varphi)&=\frac{\Lambda}{1- {\cinf}^2}\frac{\left(1-{\cinf}^2\right)^2\sinh^2\left(\xi\varphi\right)+2(1-{\cinf}^2)-1}{1+\left(1-{\cinf}^2\right)\sinh^2\left(\xi\varphi\right)} \;,
\end{align}
with $\xi=\sqrt{3\Lambda/[4(1-{\cinf}^2)M^{4}]}$. Hence, the sound speed takes the parametric form
\begin{equation}
c_s(a)=\sqrt{\frac{{\ol \cinf}^2}{\ol+(1-{\cinf}^2)\odm a^{-3}}}\label{c_s-udm},
\end{equation}
and it is easy to see that the parameter $\cinf$ represents the value of the speed of sound when $a\rightarrow\infty$. Moreover, when $a\to0$, $c_s\to0$.

\begin{figure}
\centering
\includegraphics[width=0.45\textwidth]{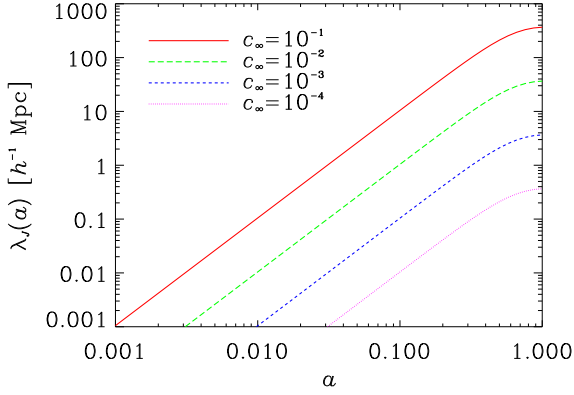}
\caption{Sound horizon $\lambda_J(a)$ for $\cinf=10^{-4},10^{-3},10^{-2},10^{-1}$ from bottom to top.}\label{lambdaJ}
\end{figure}

In UDM models the fluid which triggers the accelerated expansion at late times is also the one which has to cluster in order to produce the structures we see today. Thus, from recombination to the present epoch, the energy density of the Universe is dominated by a single dark fluid, and therefore the gravitational potential evolution is determined by the background and perturbation evolution of this fluid alone. As a result, the general trend is that the possible appearance of a sound speed significantly different from zero at late times corresponds to the appearance of a Jeans length \citep{Bertacca:2007cv}
\begin{equation}
\lambda_J(a)=\sqrt{\left|\frac{\theta}{\theta''}\right|}c_s(a)\label{jeans}
\end{equation}
below which the dark fluid does not cluster any more, causing a strong evolution in time of the gravitational potential. In Fig.~\ref{lambdaJ} we show $\lambda_J(a)$, the sound horizon, for different values of $\cinf$.

\section{Weak lensing on the celestial sphere}\label{3dlensing}
In the linear r\'egime, corresponding to the Born approximation, where the lensing effects are evaluated on the null-geodesic of the unperturbed (unlensed) photon \citep{Hu:2000ee,Bartelmann:1999yn}, it is possible to relate the weak lensing potential $\phi$ for a given source at a 3D position in comoving space $\mathbf x=(\chi,\hat{\mathbf n})$ to the Newtonian potential $\Phi(\mathbf x)$ via
\begin{equation}
\phi(\mathbf x)=\int_0^\chi\!\!\de\chi'\,\frac{W(\chi')}{\chi'}\Phi(\chi',\hat{\mathbf n})\label{phi}
\end{equation}
where
\begin{equation}
W(\chi')=-2\int_{\chi'}^\infty\de\chi\,\frac{\chi-\chi'}{\chi}n(\chi)\label{W(z)}
\end{equation}
is the weight function of weak lensing, with $n\left[\chi(z)\right]$ representing the redshift distribution of sources, for which $\int\!\!\de\chi\,n(\chi)=1$ holds, and $\chi(z)$ being the radial comoving distance, such that
\begin{equation}
\frac{1}{H(z)}=\frac{\de\chi(z)}{\de z}.
\end{equation}

Spin-weighted spherical harmonics and spherical Bessel functions are a very natural expansion for weak lensing observables, such as the potential $\phi(\mathbf x)$ \citep{Heavens:2003jx,Castro:2005bg}. Since cosmic shear depends on the Newtonian potential, the use of this basis allows one to relate the expansion of the shear field to the expansion of the mass density field. The properties of the latter depend in a calculable way on cosmological parameters, so this opens up the possibility of using 3D weak shear to estimate these quantities.

In the flat-sky approximation, the weak lensing potential (\ref{phi}) reads
\begin{equation}
\phi(k,\bmath{\ell})=\sqrt{\frac{2}{\pi}}\int\!\!\de^3x\,\phi(\mathbf x)kj_\ell\left(k\chi\right)e^{-i\bmath{\ell}\cdot\hat{\mathbf n}},
\end{equation}
where $\ell=|\bmath\ell|$ is a 2D angular wavenumber, $k$ a radial wavenumber and $j_\ell(k\chi)$ a spherical Bessel function of order $\ell$. The covariances of these coefficients define the power spectrum of the weak lensing potential via
\begin{equation}
\langle\phi(k,\bmath{\ell})\phi^\ast(k',\bmath{\ell}')\rangle={(2\pi)}^2\delta_D(\bmath{\ell}-\bmath{\ell}')C^{\phi\phi}(k,k';\ell),
\end{equation}
where $\delta_D$ is the Dirac delta.

\subsection{The 3D shear field}\label{3dshear}
In this paper we are interested in the information brought by the cosmic shear. We now introduce a distortion tensor \citep{Kaiser:1996tp,Bartelmann:1999yn}
\begin{equation}
\phi_{,ij}(\mathbf x)=\int_0^\chi\!\!\de\chi'\,\chi'W(\chi')\Phi_{,ij}(\chi',\hat{\mathbf n})\label{phi,ij},
\end{equation}
where commas denote derivatives w.r.t. directions perpendicular to the line of sight. The trace of the distortion tensor represents the convergence
\begin{equation}
\kappa(\mathbf x)=\frac{1}{2}\left(\phi_{,11}(\mathbf x)+\phi_{,22}(\mathbf x)\right)
\end{equation}
and, defining $\gamma_1(\mathbf x)=\frac{1}{2}\left(\phi_{,11}(\mathbf x)-\phi_{,22}(\mathbf x)\right)$ and $\gamma_2(\mathbf x)=\phi_{,12}(\mathbf x)$, the linear combination
\begin{equation}
\gamma(\mathbf x)=\gamma_1(\mathbf x)+i\gamma_2(\mathbf x)
\end{equation}
is the differential stretching, or shear. \citet{Castro:2005bg} have shown that the complex shear is the second ``edth'' derivative of the weak lensing potential
\begin{equation}
\gamma(\mathbf x)=\frac{1}{2}\eth\eth\phi(\mathbf x),
\end{equation}
where, in Cartesian coordinates $\{x,\,y\}$, $\eth=\partial_x+i\partial_y$.

We can now express the power spectrum of the 3D cosmic shear as a function of the gravitational potential via
\begin{equation}
C^{\gamma\gamma}(k_1,k_2;\ell)=\frac{\ell^4}{\pi^2}\int\!\!\de k\,k^2I^{\Phi}_\ell(k_1,k)I^{\Phi}_\ell(k_2,k)P^{\Phi}(k,0),\label{Cgammagamma}
\end{equation}
where $P^{\Phi}(k,z)$ is the Newtonian potential power spectrum and, for a generic field $X$, we have defined
\begin{equation}
I^X_\ell(k_i,k)=\int\!\!\de\chi\,\frac{X_k(\chi)}{X_k(0)}W(\chi)j_\ell\left(k_i\chi\right).\label{Iphi}
\end{equation}

\section{Fisher matrix analysis}\label{fisher}
Cosmological parameters influence the shear in a number of ways: the matter power spectrum $P^\delta(k,z)$ is dependent on $\om$, $h$ and the linear amplitude $\sigma_8$. The linear power spectrum is dependent on the growth rate, which also has some sensitivity to the parameter of the $\Lambda$-like component equation of state $w_\Lambda=p_\Lambda/\rho_\Lambda$. It is well know that the speed of sound (Eq.~\ref{c_s}) is strictly related to $w_\Lambda(z)$, and it also affects the $\chi(z)$ relation and hence the angular diameter distance $\sin_K\left[\chi(z)\right]$. These parameters $\{\vartheta_\alpha\}$ may be estimated from the data using likelihood methods. Assuming uniform priors for the parameters, the maximum a posteriori probability for the parameters is given by the maximum likelihood solution. We use a Gaussian likelihood
\begin{equation}
2\ln L=-\mathrm{Tr}\left[\ln\mathbfss C-\mathbfss C^{-1}\mathbfss D\right],
\end{equation}
where $\mathbfss C=\langle(\mathbf d-\mathbf d^\mathrm{th})(\mathbf d-\mathbf d^\mathrm{th})^T\rangle$ is the covariance matrix and $\mathbfss D=(\mathbf d-\mathbf d^\mathrm{th})(\mathbf d-\mathbf d^\mathrm{th})^T$ is the data matrix, with $\mathbf d$ the data vector and $\mathbf d^\mathrm{th}$ the theoretical mean vector.

The expected errors on the parameters can be estimated with the Fisher information matrix \citep{Fisher:1935,Jungman:1995bz,Tegmark:1996bz}. This has the great advantage that different observational strategies can be analysed and this can be very valuable for experimental design. The Fisher matrix gives the best errors to expect, and should be accurate if the likelihood surface near the peak is adequately approximated by a multivariate Gaussian.

The Fisher matrix is the expectation value of the second derivative of the $\ln L$ w.r.t. the parameters $\{\vartheta_\alpha\}$:
\begin{equation}
\mathbfss F_{\alpha\beta}=-\left\langle\frac{\partial^2\ln L}{\partial\vartheta_\alpha\partial\vartheta_\beta}\right\rangle\label{fisherm}
\end{equation}
and the marginal error on parameter $\vartheta_\alpha$ is $\left[\left(\mathbfss F^{-1}\right)_{\alpha\alpha}\right]^{\frac{1}{2}}$. If the means of the data are fixed, the Fisher matrix can be calculated from the covariance matrix and its derivatives \citep{Tegmark:1996bz} by
\begin{equation}
\mathbfss F_{\alpha\beta}=\frac{1}{2}\mathrm{Tr}\left[\mathbfss C^{-1}\mathbfss C_{,\alpha}\mathbfss C^{-1}\mathbfss C_{,\beta}\right].
\end{equation}
For a square patch of sky, the Fourier transform leads to uncorrelated modes, provided the modes are separated by $2\pi/\Theta_\mathrm{rad}$ where $\Theta_\mathrm{rad}$ is the side of the square in radians, and the Fisher matrix is simply the sum of the Fisher matrices of each $\ell$ mode:
\begin{equation}
\mathbfss F_{\alpha\beta}=\frac{1}{2}\sum_\ell(2\ell+1)\mathrm{Tr}\left[\left(\mathbfss C^\ell\right)^{-1}{\mathbfss C^\ell}_{,\alpha}\left(\mathbfss C^\ell\right)^{-1}{\mathbfss C^\ell}_{,\beta}\right],
\end{equation}
where $\mathbfss C^\ell$ is the covariance matrix for a given $\ell$ mode.

\section{Bayesian evidence}\label{B-evidence}
In this paper we compute parameter forecasts from 3D cosmic shear for UDM models. It is important to notice that we are dealing with an alternative model with respect to the standard $\Lambda$CDM model; hence, besides determining the best-fit value (and the errors) on a set of parameters within a model, we can also ask if this particular alternative model is preferable to the standard. Model selection is in a sense a higher-level question than parameter estimation. While in estimating parameters one assumes a theoretical model within which one interprets the data, in model selection, one wants to know which theoretical framework is preferred given the data. Clearly if our alternative model has more parameters than the standard one, chi-square analysis will not be of any use, because it will always reduce if we add more degrees of freedom. From a Bayesian point of view, this involves computation of the Bayesian evidence and of the Bayes factor $B$.

We refer to the two models under examination with $M_\mathrm{UDM}$ and $M_\textrm{$\Lambda$CDM}$. We know that, in this context, $M_\textrm{$\Lambda$CDM}$ is simpler than $M_\mathrm{UDM}$ because it has one fewer parameter, i.e. $\cinf$; in the same way, it is also contained in $M_\mathrm{UDM}$, because, if  $\vartheta^\textrm{$\Lambda$CDM}_\alpha$ and $\vartheta^\mathrm{UDM}_{\alpha'}$ are the parameters of the two models (with $\alpha=1,\ldots,n$ and $\alpha'=1,\ldots,n+1$), respectively, then
\begin{equation}
\{\vartheta^\textrm{$\Lambda$CDM}_\alpha,\,\cinf\}=\{\vartheta^\mathrm{UDM}_{\alpha'}\}
\end{equation}
holds; here, $\cinf\equiv\vartheta^\mathrm{UDM}_{n+1}$.

The posterior probability for each model $M$ is given by Bayes' theorem
\begin{equation}
p(M|\mathbf d)=\frac{p(\mathbf d|M)p(M)}{p(\mathbf d)}.
\end{equation}
The Bayesian evidence is defined as the marginalisation over the parameters
\begin{equation}
p(\mathbf d|M)=\int\!\!\de^m\vartheta\,p(\mathbf d|\bvartheta,M)p(\bvartheta|M),
\end{equation}
where $\bvartheta$ is the parameter vector, whose length $m$ is $n$ for the $\Lambda$CDM model and $n+1$ for UDM models. The posterior relative probabilities of our two models given the data $\mathbf d$ and with flat priors in their parameters $p(M)=\textrm{const.}$, is then \citep{Heavens:2007ka,Heavens:2009nx}
\begin{multline}
\frac{p(M_\textrm{$\Lambda$CDM}|\mathbf d)}{p(M_\mathrm{UDM}|\mathbf d)}=\frac{p(M_\textrm{$\Lambda$CDM})}{p(M_\mathrm{UDM})}\\\times\frac{\int\!\!\de^n\vartheta^\textrm{$\Lambda$CDM}\,p(\mathbf d|\bvartheta^\textrm{$\Lambda$CDM},M_\textrm{$\Lambda$CDM})p(\bvartheta^\textrm{$\Lambda$CDM}|M_\textrm{$\Lambda$CDM})}{\int\!\!\de^{n+1}\vartheta^\mathrm{UDM}\,p(\mathbf d|\bvartheta^\mathrm{UDM},M_\mathrm{UDM})p(\bvartheta^\mathrm{UDM}|M_\mathrm{UDM})}.
\end{multline}

If we choose non-committal priors $p(M_\mathrm{UDM})=p(M_\textrm{$\Lambda$CDM})$, the posterior evidence probability reduces to the ratio of the evidences, which takes the name of the Bayes factor and in the present case reads
\begin{equation}
B\equiv\frac{\int\!\!\de^n\vartheta^\textrm{$\Lambda$CDM}\,p(\mathbf d|\bvartheta^\textrm{$\Lambda$CDM},M_\textrm{$\Lambda$CDM})p(\bvartheta^\textrm{$\Lambda$CDM}|M_\textrm{$\Lambda$CDM})}{\int\!\!\de^{n+1}\vartheta^\mathrm{UDM}\,p(\mathbf d|\bvartheta^\mathrm{UDM},M_\mathrm{UDM})p(\bvartheta^\mathrm{UDM}|M_\mathrm{UDM})}.
\end{equation}

Now, let us focus on the priors $p(\bvartheta|M)$. If we assume flat priors in each parameter, over the range $\Delta\bvartheta$, then $p(\bvartheta^\textrm{$\Lambda$CDM}|M_\textrm{$\Lambda$CDM})=\prod_\alpha\left(\Delta\vartheta^\textrm{$\Lambda$CDM}_\alpha\right)^{-1}$ and
\begin{equation}
B=\frac{\int\!\!\de^n\vartheta^\textrm{$\Lambda$CDM}\,p(\mathbf d|\bvartheta^\textrm{$\Lambda$CDM},M_\textrm{$\Lambda$CDM})}{\int\!\!\de^{n+1}\vartheta^\mathrm{UDM}\,p(\mathbf d|\bvartheta^\mathrm{UDM},M_\mathrm{UDM})}\Delta \cinf.
\end{equation}

The Bayes factor $B$ still depends on the specific dataset $\mathbf d$. For future experiments, we do not yet have the data, so we compute the expectation value of the Bayes factor, given the statistical properties of $\mathbf d$. The expectation is computed over the distribution of $\mathbf d$ for the correct model (assumed here to be $M_\mathrm{UDM}$). To do this, we make two further approximations: first we note that $B$ is a ratio, and we approximate $\langle B\rangle$ by the ratio of the expected values, rather than the expectation value of the ratio. This should be a good approximation if the likelihoods are sharply peaked.

We also make the Laplace approximation, that the expected likelihoods are given by multivariate Gaussians, i.e.,
\begin{equation}
p(\mathbf d|\bvartheta,M)=L_0e^{-\frac{1}{2}\left(\vartheta-\vartheta_0\right)_\alpha\mathbfss F_{\alpha\beta}\left(\vartheta-\vartheta_0\right)_\beta},
\end{equation}
where $\mathbfss F_{\alpha\beta}$ is the Fisher matrix, given in Eq.~(\ref{fisherm}). \citet{Heavens:2007ka} have shown that, if we assume that the posterior probability densities are small at the boundaries of the prior volume, then we can extend the integration to infinity, and the integration over the multivariate Gaussians can be easily performed. In the present case, this gives
\begin{equation}
\langle B\rangle=\frac{\sqrt{\det\mathbfss F^\mathrm{UDM}}}{\sqrt{\det\mathbfss F^\textrm{$\Lambda$CDM}}}\frac{L^\textrm{$\Lambda$CDM}_0}{L^\mathrm{UDM}_0}\frac{\Delta \cinf}{\sqrt{2\pi}}.
\end{equation}

One more subtlety has to be taken into account to compute the ratio $L^\textrm{$\Lambda$CDM}_0/L^\mathrm{UDM}_0$: if the correct underlying model is $M_\mathrm{UDM}$, in the incorrect model $M_\textrm{$\Lambda$CDM}$ the maximum of the expected likelihood will not, in general, be at the correct parameter values \citep[see][Fig.~1]{Heavens:2007ka}. The $n$ parameters of the $\Lambda$CDM model shift their values to compensate the fact that $\cinf$ is being kept fixed  at the incorrect fiducial value $\cinf=0$. With these offsets in the maximum likelihood parameters in the $\Lambda$CDM model, the Bayes factor takes the form
\begin{equation}
\langle B\rangle=\frac{\sqrt{\det\mathbfss F^\mathrm{UDM}}}{\sqrt{\det\mathbfss F^\textrm{$\Lambda$CDM}}}\frac{\Delta \cinf}{\sqrt{2\pi}}e^{-\frac{1}{2}\delta\vartheta_\alpha\mathbfss F^\mathrm{UDM}_{\alpha\beta}\delta\vartheta_\beta},\label{B}
\end{equation}
where the shifts $\delta\vartheta_\alpha$ can be computed under the assumption of a multivariate Gaussian distribution \citep{Taylor:2006aw}, and read
\begin{equation}
\delta\vartheta_\alpha=-\left[\left(\mathbfss F^\textrm{$\Lambda$CDM}\right)^{-1}\right]_{\alpha\beta}\mathbfss G^\mathrm{UDM}_{\beta,n+1}\delta \cinf,\label{shifts}
\end{equation}
with $\mathbfss G^\mathrm{UDM}_{\beta,n+1}$ a subset of the UDM Fisher matrix (a vector in the present case).

It is usual to consider the logarithm of the Bayes factor, for which the so-called ``Jeffreys' scale'' gives empirically calibrated levels of significance for the strength of evidence \citep{Jeffreys:1961}, $1<|\ln B|<2.5$ is described as ``substantial'' evidence in favour of a model, $2.5<|\ln B|<5$ is ``strong,'' and $|\ln B|>5$ is ``decisive.'' These descriptions seem too aggressive: $|\ln B|=1$ corresponds to a posterior probability for the less-favoured model which is $0.37$ of the favoured model \citep{Kass95bayesfactors}. Other authors have introduced different terminology \citep[e.g.][]{Trotta:2005ar}.

\section{Results and discussion}\label{results}
We use a fiducial cosmology with the following parameters: Hubble constant (in units of $100\,\mathrm{km\,s^{-1}\,Mpc^{-1}}$) $h=0.71$, present-day total matter density (in units of critical density) $\om\equiv\odm+\ob=0.3$, baryon contribution $\ob=0.045$, cosmological constant contribution $\ol=0.7$, spectral index $n_s=1$, linear amplitude (within a sphere of radius $8\,h^{-1}\,\mathrm{Mpc}$) $\sigma_8=0.8$.

In Section~\ref{matterpowerspectrum} we compute the predicted matter power spectrum for UDM models, with a comparison to $\Lambda$CDM. In Section~\ref{signal} the 3D shear matrix $C^{\gamma\gamma}(k_1,k_2;\ell)$ is shown. In Section~\ref{estimation} we present the parameter forecasts, and in Section~\ref{selection} we show the expected Bayesian evidence for UDM models over the $\Lambda$CDM model.

\subsection{The matter power spectrum}\label{matterpowerspectrum}
Our class of UDM models allows the value $w=-1$ for $a\to\infty$. In other words they admit an effective cosmological constant energy density at late times. Therefore, in order to compare the predictions of our UDM model with observational data, we follow the same prescription used by \citet{Piattella:2009kt}, where the density contrast of the clustering fluid is
\begin{equation}
\delta\equiv\frac{\delta\rho_m}{\rho_m}=\frac{\rho_\mathrm{DM}\delta_\mathrm{UDM}+\rho_b\delta_b}{\rho_m},
\end{equation}
where $\delta_b$ and $\delta_\mathrm{UDM}$ are the baryon and the scalar field density contrasts, respectively, and we emphasise that $\rho_\mathrm{DM}=\rho_\mathrm{UDM}-\rho_\Lambda$ is the only component of the scalar field density which clusters.

\subsubsection{Linear r\'egime}
The today matter power spectrum $P(k)\equiv P^\delta\left(k,z=0\right)$ is the present value of the Fourier transform of the density perturbation correlation function. To construct $P(k)$ in the $\Lambda$CDM model, we need the growth factor $D(z)=\delta(\mathbf x,z)/\delta(\mathbf x,z=0)$ on linear scales (i.e. in absence of free-streaming) and the transfer function $T(k)$, that describes the evolution of perturbations through the epochs of horizon crossing and radiaton-matter transition. Here, we use the transfer function suggested by \citet{Eisenstein:1997jh}, which, with an accurate, general fitting formula, calculates the power spectrum as a function of the cosmological parameters quite efficiently. \citet{Eisenstein:1997jh} show that baryons are effective at suppressing power on small scales compared to DM-only models. Moreover, the small-scale limit of this transfer function can be calculated analytically as a function of the cosmological parameters \citep{Hu:1997vi}. Hence, we can write the matter power spectrum as
\begin{equation}
P(k)=2\pi^2{\delta_H}^2\left(\frac{k}{\ho^3}\right)^{n_s}T^2(k)\left[\frac{D(z)}{D(z=0)}\right]^2;
\end{equation}
here, $\delta_H$ is a normalisation.

To obtain $P(k)$ in UDM models, it is useful to remember that the class of UDM models we use here is constructed to have the same properties of the $\Lambda$CDM model in the early Universe; in Eq.~(\ref{eq-Mukhanov:2005sc-lcdm}), which describes the time evolution of Fourier modes of the Newtonian potential $\Phi_\mathbf{k}(a)$, we thus set the same initial conditions for both the UDM and the $\Lambda$CDM potentials. Gravity is GR, so we can use the Poisson equation
\begin{equation}
\Phi_\mathbf{k}(a)=-\frac{3}{2}\om{\ho}^2\frac{\delta_\mathbf k(a)}{k^2a}\label{poisson},
\end{equation}
which relates $\Phi_\mathbf{k}(a)$ to the matter power spectrum via
\begin{equation}
\langle\delta_\mathbf{k}(a){\delta_{\mathbf k'}}^\ast(a)\rangle=\left(2\pi\right)^3\delta_D\left(\mathbf k-\mathbf k'\right)P^\delta(k,a).
\end{equation}
Clearly, if we solve Eq.~(\ref{eq-Mukhanov:2005sc-lcdm}) with $c_s=0$, we obtain the standard $\Lambda$CDM matter power spectrum.

\begin{figure}
\centering
\includegraphics[width=0.5\textwidth]{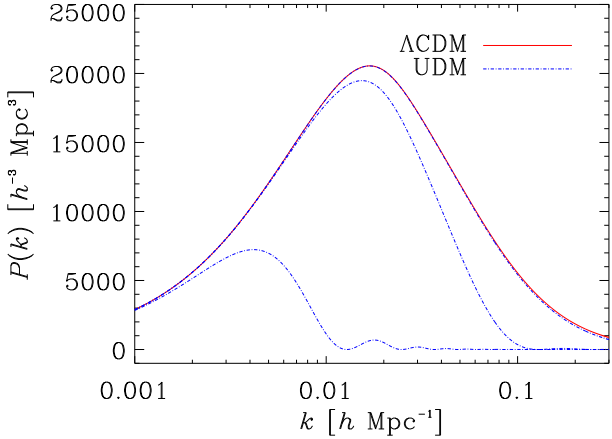}
\caption{Matter power spectra $P(k)\equiv P^\delta(k,0)$ for $\Lambda$CDM (solid) and UDM (dot-dashed), with $\cinf=10^{-3},10^{-2},10^{-1}$ from top to bottom.}\label{matter_power_spetrum}
\end{figure}

Fig.~\ref{matter_power_spetrum} shows the matter power spectrum $P(k)$ for $\Lambda$CDM and UDM models, for a number of values of $\cinf$. By increasing the sound speed, the potential starts to decay earlier in time, oscillating around zero afterwards \citep{Camera:2009uz}; at large scales, if $\cinf$ is small enough, these UDM models reproduce the $\Lambda$CDM model. This feature reflects the dependence of the gravitational potential on the effective Jeans length $\lambda_J(a)$. It is easy to see that if $\cinf\lesssim10^{-3}$ the perturbations of the UDM reproduce the behaviour of the concordance model within the linear r\'egime (the UDM curve for $\cinf=10^{-3}$ is virtually on top of the $\Lambda$CDM one). Instead, a larger sound speed inhibits structure formation earlier in time, thus we observe less power on small scales; in this case, the consequence of the oscillatory feature of the gravitational potential, due to the non-negligible speed of sound, can be clearly seen.

In principle, the large-scale distribution of galaxies could constrain the value of $\cinf$. However, the shape of the power spectrum also depends on the normalisation $\sigma_8$ and the spectral index $n_s$: therefore, for a given $\cinf$ as large as $10^{-2}$, an appropriate choice of $\sigma_8$ and $n_s$ can provide a power spectrum in agreement with observations, at least on scales where non-linear effects are not dominant. In addition, in UDM models it is still unclear how the galaxy distribution is biased against the gravitational potential of the scalar field on small scales. Therefore the large-scale distribution of galaxies does not appear to be the best tool to constrain this family of UDM models. On the contrary, a weak lensing analysis can constrain the matter power spectrum without a fine-tuning of either $\sigma_8$ or the galaxy bias.

\subsubsection{Non-linear r\'egime}
For wavenumbers $k>k_\mathrm{nl}\simeq0.2\,h\,\mathrm{Mpc}^{-1}$, non-linear contributions to the evolution of the Newtonian potential (i.e. to matter overdensities) become important. In the $\Lambda$CDM model, the gravitational potential satisfies Eq.~(\ref{eq-Mukhanov:2005sc-lcdm}), but in this case $c_s$ is the sound speed of the hydrodynamical fluid, and therefore can be set equal to zero in the matter-dominated epoch. For $c_s=0$, Eq.~(\ref{eq-Mukhanov:2005sc-lcdm}) has an analytic solution \citep{Hu:1998tj,Hu:2001fb,Mukhanov:2005sc,Bertacca:2007cv}
\begin{equation}
\Phi_\mathbf{k}(a)=A_\mathbf{k}\left(1-\frac{H(a)}{a}\int_0^a\!\!\frac{\de a'}{H(a')}\right),\label{longwave}
\end{equation}
where the constant of integration is $A_\mathbf{k}=\Phi_\mathbf{k}(0)T(k)$, with $T(k)$ the matter transfer function and $\Phi_\mathbf{k}(0)$ the large-scale potential during the radiation-dominated era.

To perform further calculations on a wider range of scales than that allowed by linear theory, we will use the \citet{Smith:2002dz} non-linear fitting formul\ae~ for $P(k)$ in the $\Lambda$CDM model. However, currently there is no linear-to-non-linear mapping in UDM models. Nevertheless, as we have seen, differences between the $\Lambda$CDM and UDM models arise at scales smaller than the sound horizon. With a cross-over wavenumber $k\simeq1/\lambda_J$, if the sound speed is small enough to guarantee that $\lambda_J$ is well within the non-linear regime we can assume that the non-linear evolution of the UDM power spectrum will be similar to the $\Lambda$CDM one. A deeper knowledge on this aspect will be the next step of the development of UDM models and has to be explored in future work.

\subsection{The 3D shear signal}\label{signal}
For a $20,000\,\mathrm{deg}^2$ Euclid-like survey \citep{Cimatti:2009is,Refregier:2010ss}, we assume that the source distribution over redshifts has the form \citep{1994MNRAS.270..245S}
\begin{equation}
\bar n(z)\propto z^2e^{-\left(\frac{z}{z_0}\right)^{1.5}},
\end{equation}
where $z_0=z_m/1.4$, and $z_m=0.8$ is the median redshift of the survey. The source number density with photometric redshift and shape estimates is $35$ per square arcminute. We also assume that the photometric redshift errors are Gaussian, with a dispersion given by $\sigma(z)=0.05(1+z)$.

In order to avoid the high-wavenumber r\'egime where the fitting formul\ae~ of \citet{Smith:2002dz} may be unreliable, or where baryonic effects might alter the power spectrum ($k>10\,h\,\mathrm{Mpc}^{-1}$; \citealt{White:2004kv,Zhan:2004wq}), we do not analyse modes with $k>1.5\,\mathrm{Mpc}^{-1}$. Note that the non-local nature of gravitational lensing does mix modes to some degree, but these modes are sufficiently far from the uncertain highly non-linear r\'egime that this is not a concern \citep{Castro:2005bg}. We include angular modes as small as each survey will allow, and analyse up to $\ell_\mathrm{max}=5000$ (but note the wavenumber cut).

\begin{figure*}
\centering
\includegraphics[width=\textwidth]{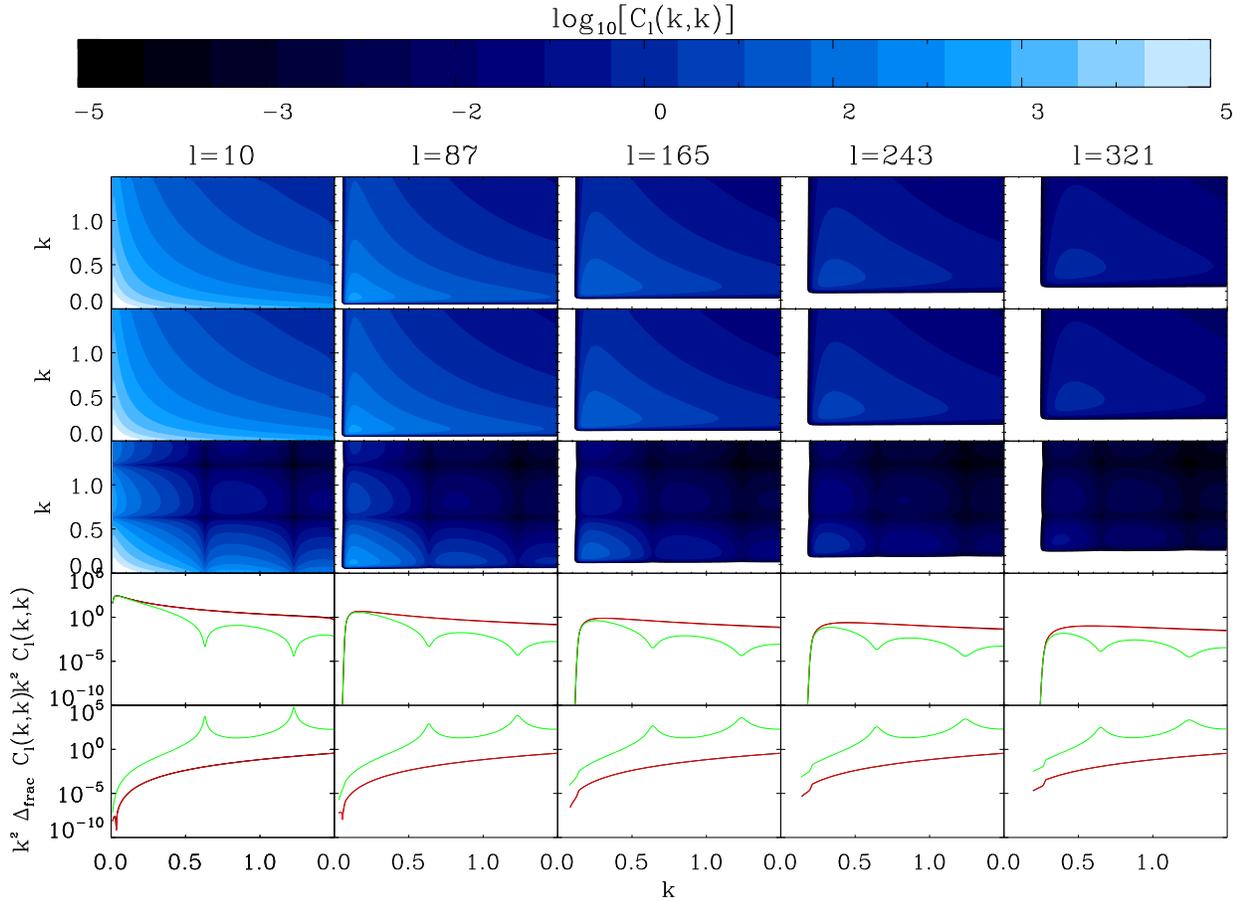}
\caption{The 3D shear matrix $\log_{10}C^{\gamma\gamma}(k_1,k_2;\ell)$ for five values of $\ell$ in (blue)gray-scale. In the first row we show the $\Lambda$CDM signal, while in the second and third rows we present the UDM signal for $\cinf=1.0\,\cdot10^{-3}$ and $\cinf=5.4\cdot10^{-3}$, respectively. The fourth row shows the diagonal elements $k^2C^{\gamma\gamma}(k,k;\ell)$, and each curve, from top to bottom, refers to the corresponding matrix above. The $\Lambda$CDM curve is virtually on top of the small-$\cinf$ UDM curve. The fifth row shows the fractional error.}\label{shearsignal}
\end{figure*}

In Fig.~\ref{shearsignal} we present the 3D shear matrix $C^{\gamma\gamma}(k_1,k_2;\ell)$. The first three rows show $\log_{10}C^{\gamma\gamma}(k_1,k_2;\ell)$ for the $\Lambda$CDM model and for a UDM model with $\cinf=5\,\cdot10^{-4}$ and $\cinf=5\cdot10^{-3}$ (respectively) in the $(k_1,k_2)$-plane in blue(gray)-scale for a number of values of $\ell$. In the fourth row we present the diagonal elements $k^2C^{\gamma\gamma}(k,k;\ell)$ of the 3D shear matrix, where the upper (green) curve refers to the smaller speed of sound and the lower (green) curve to the greater $\cinf$; the $\Lambda$CDM (red) curve is virtually on top of the small-$\cinf$ UDM curve. Finally, in the bottom row we show $k^2$ times the ratio of the diagonal elements $C^{\gamma\gamma}(k,k;\ell)$ of UDM models over the $\Lambda$CDM model.

The oscillatory features of the UDM gravitational potential \citep{Camera:2009uz}, whose power spectrum enters the shear via Eq.~(\ref{Cgammagamma}), can be clearly seen in the shear signal of the UDM model with $\cinf=4\cdot10^{-3}$. The bumps in the diagonal signal can be easily understood by looking at the $\log_{10}C^{\gamma\gamma}(k_1,k_2;\ell)$ plot, where it is interesting to notice how the oscillations take place along any direction, with the obvious symmetry along the $k_1$- and $k_2$-axes. Instead, as we have noticed in Fig.~\ref{matter_power_spetrum}, when the sound speed is small enough we do not see any oscillations and the matter power spectrum of UDM models is in agreement with $\Lambda$CDM. This agreement holds even at non-linear scales $k\gtrsim0.2\,h\,\mathrm{Mpc}^{-1}$.

Beyond the oscillations, these signals, expected for two different values of $\cinf$, show us the effect of the effective Jeans length of the gravitational potential. In fact, The Newtonian potential in UDM models behaves like $\Lambda$CDM at scales much larger than $\lambda_J(a)$ (Eq.~\ref{jeans}), while at smaller scales it starts to decay and oscillate. Hence, at high values of $\ell$ and $k$, which correspond to small angular and physical scales, respectively, the signal of weak lensing observables, like cosmic shear, shows the decay of the gravitational potential.

Although the UDM signal for $\cinf=5\,\cdot10^{-4}$ appears to be in agreement with the $\Lambda$CDM signal (fourth row of Fig. \ref{shearsignal}), their fractional difference shown in the fifth row is still of order unit at $k\gtrsim1\,h\,\mathrm{Mpc}^{-1}$ and is not negligible. In fact, we will see below in Section~\ref{selection}, that this low value of $\cinf$ still yields a Bayesian evidence which indicates a statistically very large difference between this UDM model and $\Lambda$CDM.

Finally, in Fig.~\ref{shearsignal}, we can also notice that, the higher the value of $\ell$, the smaller the physical scales are those which contribute to the shear signal. This effect is due to the approximate Bessel function inequality, $\ell\leq k\chi$, in Eq.~(\ref{Iphi}). As the $\ell$ value increases the diagonal terms of the covariance matrix do not become significant until $k\chi_\mathrm{max}\sim\ell$, where $\chi_\mathrm{max}\equiv\chi(z_\mathrm{max})$ is the upper limit imposed on the integration over the radial comoving distance.

\subsection{Estimation of cosmological parameters}\label{estimation}
Once having introduced the method (Section~\ref{fisher}) and the survey design formalism (Section~\ref{signal}), now we show cosmological parameter forecasts for such a survey and we explore the variation in the marginal errors with changes in the sound speed parameter $\cinf$.

By using the Fisher matrix analysis outlined in \citet{Taylor:2006aw}, we calculate predicted Fisher matrices and parameter constraints for a $20,000$ square-degree Euclid-like survey. In all Fisher matrix calculations we use a seven-parameter cosmological set $\{\om=\odm+\ob,\,\ob,\,h,\,\ol,\,\sigma_8,\,n_s,\,\cinf\}$ with fiducial values $\{0.3,\,0.045,\,0.71,\,0.7,\,0.8,\,1.0\}$ for the first six. The Fisher matrix is sensitive to $\cinf$, so we compute the evidences at twenty $\cinf$ fiducial values from $5\cdot10^{-4}$ to $5\cdot10^{-2}$. We find that the Fisher matrices are unstable for $\cinf\lesssim10^{-3}$. This is because, when the sound speed is small, the UDM 3D shear signal is virtually indistinguishable from that of \lcdm, and the numerical derivatives w.r.t. $\cinf$ thus become unreliable.

\begin{figure*}
\centering
\includegraphics[width=\textwidth,trim=10mm 20mm 20mm 0mm,clip]{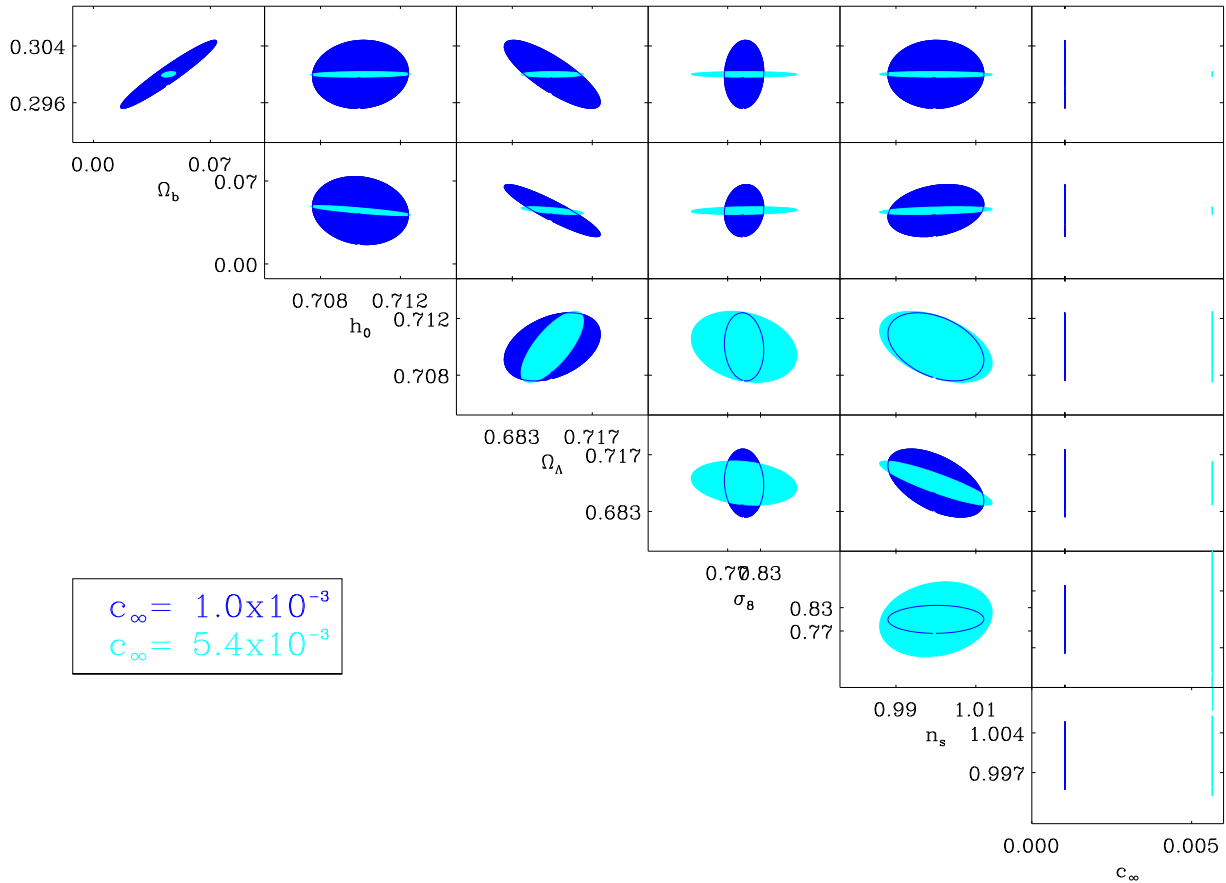}
\caption{Expected marginal errors on UDM model cosmological parameters from a $20,000\,\mathrm{deg}^2$ Euclid-like survey with a median redshift $z_m=0.8$. Ellipses show the $1\sigma$ errors for two parameters ($68\%$ confidence regions), marginalised over all the other parameters. Dark(light) ellipses refer to a UDM model with $\cinf=1.0\cdot10^{-3}$($\cinf=5.4\cdot10^{-3}$).}\label{fisher_plot}
\end{figure*}

Fig.~\ref{fisher_plot} shows the Fisher matrix elements marginalised over all other parameters. In dark blue(gray) we present the results for a UDM model with $\cinf=1.0\cdot10^{-3}$ and in light blue(gray) for $\cinf=5.4\cdot10^{-3}$. Notice that results are shown for universes which are not necessarily flat. In non-flat geometries, the spherical Bessel functions $j_\ell(k\chi)$ should be replaced by ultraspherical Bessel functions $\Phi^\ell_\beta(y)$ \citep{Heavens:2006uk}. For the case considered here $\ell\gg1$ and $k\gg\left(\textrm{curvature scale}\right)^{-1}$, then $\Phi^\ell_\beta(y)\to j_\ell(k\chi)$ \citep{Abbott:1986ct,Zaldarriaga:1999ep}. The expansion used is not ideal for curved universes, but it should however be an adequate approximation given current constraints on flatness \citep[e.g.][]{Larson:2010gs}.

The Fisher constraints for lensing are large enough that for some parameters $(\sigma_8,\,\Omega_b)$ the $1\sigma$ confidence region has an unphysical lower bound. We note that this is a symptom of the Fisher matrices Gaussian approximation. \citet{Taylor:2010pi} address this concern by suggesting a semi-analytic approach that only assumes Gaussianity in particular parameter directions; we leave an implementation of this type of parameter error prediction, or a more sophisticated likelihood parameter search for future investigation.

Before starting the interpretation of such results, it is important to underline that what deeply affects the matter power spectrum in UDM models, and thus the lensing signal, is the presence of an effective Jeans length for the Newtonian potential. Let us focus on Eq.~(\ref{eq-Mukhanov:2005sc-lcdm}): we can consider the asymptotic solutions, i.e. long wavelength and short wavelength perturbations, depending on whether $k\ll1/\lambda_J$ or $k\gg1/\lambda_J$, respectively. In the former case, the term in Eq.~(\ref{eq-Mukhanov:2005sc-lcdm}) involving the speed of sound of the scalar field is negligible, therefore the solution is formally the same that in the $\Lambda$CDM model (Eq.~\ref{longwave}), and the Fourier modes $\Phi_k(a)$ read \citep{Bertacca:2007cv}
\begin{equation}
\Phi_{k\ll1/\lambda_J}(a)\propto\left[1-\frac{H(a)}{a}\int_0^a\!\!\frac{\de a'}{H(a')}\right]\qquad(k\ll1/\lambda_J);
\end{equation}
instead, in the opposite r\'egime we have
\begin{equation}
\Phi_{k\gg1/\lambda_J}(a)\propto\frac{1}{\sqrt{c_s(a)}}\cos\left[k\int_0^a\!\!\de a'\,\frac{c_s(a')}{{a'}^2H(a')}\right]\;(k\gg1/\lambda_J).
\end{equation}
This means that what enters in the oscillatory dynamics is not only $\cinf$, which however plays an important role, but also $\odm$ and $\ol$, as described in Eq.~(\ref{c_s-udm}). Therefore, the links which connect the expected marginal errors in Fig.~\ref{fisher_plot} with the corresponding fiducial $\cinf$ are not quite straightforward. Moreover, we find that the Fisher matrix is rather sensitive to $\cinf$. The errors we find on the sound speed parameter are almost constant, and go from $\Delta \cinf=3.0\cdot10^{-5}$, for the fiducial value $\cinf=1.0\cdot10^{-3}$, to $\Delta \cinf=2.6\cdot10^{-5},$ when $\cinf=1.2\cdot10^{-2}$.

It is already well known that weak lensing can tightly constrain the $(\Omega_m,\,\sigma_8)$-plane, using standard cosmic shear techniques \citep[see][]{Brown:2002wt,2006A&A...452...51S}, and 3D weak lensing constrains $\sigma_8$ in the same way by measuring the overall normalisation of the matter power spectrum. The expected marginal errors on $\om$ and $\sigma_8$ are in fact very promising, particularly in the perspective of combining the cosmic shear data with other cosmological observables, i.e. CMB or SNeIa \citep{Heavens:2006uk}. However, the presence of a sound speed can be mimicked in the power spectrum, at least in the non-linear r\'egime, by an accurate tuning of some parameter values, on top of all $\sigma_8$ and $n_s$ \citep{Camera:2009uz}. This is why the ellipses of those parameters get worse for larger values of $\cinf$.

In UDM models, there is another aspect which is particularly interesting to notice: we are able to lift the degeneracy between $\om$ and $\ob$ without using early-Universe data. That is because $\odm$ and $\ob$ enter in the growth of structures in two different ways. The expansion history of the Universe takes into account only their joint effect, through $\om$, whereas the speed of sound is determined by $\odm$ alone. In fact we have to keep in mind that in UDM models there is a scalar field which mimics both DM and $\Lambda$, but it still has proper dynamics different from that of its respective in the $\Lambda$CDM model.

\subsection{Model selection}\label{selection}
In Section~\ref{B-evidence} we showed how the Bayes factor can be used to determine which model is favoured by the data. By using the Fisher matrix formalism for a Euclid-like survey, we compute the Bayes factor $B$ for UDM models over the standard $\Lambda$CDM cosmology. We fix flat prior $\Delta \cinf=1$.
\begin{figure}
\centering
\includegraphics[width=0.5\textwidth,trim=15mm 0mm 0mm 0mm,clip]{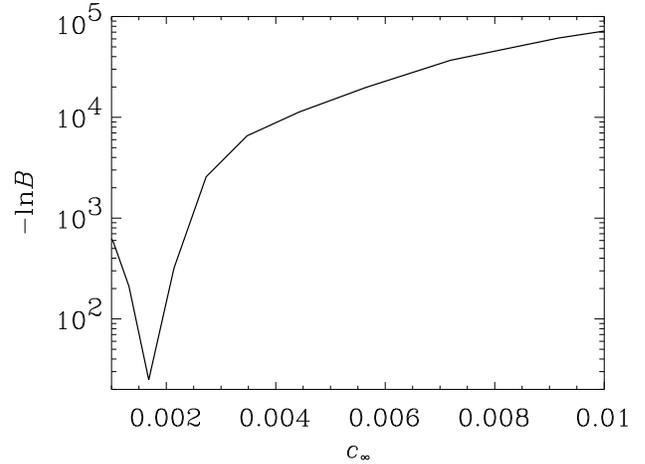}
\caption{Bayes factor $-\ln B$ for UDM models over the standard $\Lambda$CDM model as a function of the sound speed parameter $\cinf$.}\label{evidence}
\end{figure}

The large values of $-\ln B$ derive from the large deviations $\delta\vartheta_\alpha$ in Eq.~(\ref{shifts}) which yield an extremely small exponential. On turn, the deviations $\delta\vartheta_\alpha$ are large because, as shown in the right-most column of Fig.~\ref{fisher_plot}, (i) the ellipsoidal confidence regions are narrow, and (ii) they are almost vertical; in other words, the $\Lambda$CDM parameters that one would derive if living in a universe with a non-null $\cinf$ would be largely biased.

We conclude that, if UDM is the correct model, there would be large evidence for UDM models over $\Lambda$CDM for values of $\cinf\gtrsim10^{-3}$. However, if $\cinf$ is so small that the UDM peculiar features in the matter power spectrum only appear at $k\gg1\,h\,\mathrm{Mpc}^{-1}$, namely on galactic or smaller scales, in principle, we might be unable to distinguish UDM from $\Lambda$CDM, unless the non-linear dynamics and/or the effects of the baryonic physics on the DM-like dynamics of the scalar field are largely different from what we expect in $\Lambda$CDM.

\section{Conclusions}\label{conclusions}
In this work, we calculate the expected error forecasts for a $20,000$ square degree survey with median redshift $z_m=0.8$ such as Euclid \citep{Cimatti:2009is,Refregier:2010ss} in the framework of unified models of DM and DE (UDM models). We focus on those UDM models which are able to reproduce the same Hubble parameter as in the $\Lambda$CDM model \citep{Bertacca:2007cv,Bertacca:2008uf}. In these UDM models, beyond standard matter and radiation, there is only one exotic component, a classical scalar field with a non-canonical kinetic term in its Lagrangian, that during the structure formation behaves like DM, while at the present time contributes to the total energy density of the Universe like a cosmological constant $\Lambda$.

In order to avoid the strong integrated Sachs-Wolfe effect which typically plagues UDM models, we follow the technique outlined by \citet{Bertacca:2008uf}, that allows one to construct a UDM model in which the sound speed is small enough to let the cosmological structures grow and reproduce the LSS we see today. This can be achieved by parameterising the sound speed with its value at late times, $\cinf$.

An effect of the presence of a non-negligible speed of sound of the UDM scalar field is the emerging of an effective time-dependent Jeans length $\lambda_J(a)$ of the gravitational potential. It causes a strong suppression, followed by oscillations, of the Fourier modes $\Phi_{\mathbf k}(a)$ with $k\equiv|\mathbf k|>1/\lambda_J$. This reflects on the predicted lensing signal, because the latter is an integrated effect of the potential wells of the LSS over the path that the photons travel from the sources to the observer.

We calculate the 3D shear matrix $C^{\gamma\gamma}(k_1,k_2;\ell)$ in the flat-sky approximation for a large number of values of $\cinf$. In agreement with \citet{Camera:2009uz}, we see that, whilst the agreement with the $\Lambda$CDM model is good for small values of $\cinf$, when one increases the sound speed parameter, the lensing signal appears more suppressed at small scales, and moreover the 3D shear matrix does show bumps related to the oscillations of the gravitational potential.

We also compute the Fisher matrix for a Euclid-like survey. It has been shown that 3D lensing is a powerful tool in constraining cosmological parameters \citep[e.g.][]{Castro:2005bg}, and \citet{Heavens:2006uk} have demonstrated that it is particularly useful in unveiling the properties of the dark components of the Universe. By using a seven-parameter cosmological set $\{\om=\odm+\ob,\,\ob,\,h,\,\ol,\,\sigma_8,\,n_s,\,\cinf\}$, with one fiducial value for each parameter, except for $\cinf$, for which we use twenty values in the range $5\cdot10^{-4}\ldots5\cdot10^{-2}$, we obtain the expected marginal errors. However, the $\cinf$ Fisher matrix elements are unstable in the parameter range $\cinf\lesssim10^{-3}$, because the UDM signal is degenerate with respect to \lcdm. Therefore, we restrict our analysis by considering only sound speed fiducial values larger than $\sim10^{-3}$. We get minimal errors that go from $\Delta \cinf=3.0\cdot10^{-5}$, for the fiducial value $\cinf=1.0\cdot10^{-3}$, to $\Delta \cinf=2.6\cdot10^{-5},$ when $\cinf=1.2\cdot10^{-2}$.

In the case of UDM models, 3D lensing is revealed to be even more useful for estimating cosmological parameters, because since it encodes information from both the geometry and the dynamics of the Universe, it can lift the usual degeneracy between the DM and the baryon fractions, $\odm$ and $\ob$. This is because in the Hubble parameter, which determines the background evolution of the geometry of the Universe, both $\odm$ and $\ob$ enter in the usual way, through the total matter fraction $\om$. On the other side, the speed of sound, which affects the structure formation, and thus the dynamics of the Universe, is sensitive only on the DM-like behaviour of the scalar field, since for baryons $c_s=0$ holds.

Finally, we compute the Bayesian expected evidence \citep[e.g.][]{Trotta:2005ar} for UDM models over the $\Lambda$CDM model as a function of the sound speed parameter $\cinf$. The expected evidence clearly shows that the survey data would unquestionably favour UDM models over the standard $\Lambda$CDM model, if its sound speed parameter exceed $\sim10^{-3}$.

\section*{Acknowledgments}
We thank the referee for her/his useful comments which contributed to remove some ambiguities in the presentation of our results. SC and AD gratefully acknowledge partial support from the INFN grant PD51. SC acknowledges Research Grants funded jointly by Ministero dell'Istruzione, dell'Universit\`a e della Ricerca (MIUR), by Universit\`a di Torino and by Istituto Nazionale di Fisica Nucleare within the {\sl Astroparticle Physics Project} (MIUR contract number: PRIN 2008NR3EBK). SC also acknowledges partial support from the Institute for Astronomy, University of Edinburgh and thanks it for the hospitality. TDK is supported by the STFC Rolling Grant number RA0888. DB would like to acknowledge the ICG Portsmouth for the hospitality during the development of this project and the ``Fondazione Ing. Aldo Gini" for support. DB research has been partly supported by ASI contract I/016/07/0 ``COFIS".

\bibliographystyle{mn2e}
\bibliography{/home/camera/Documenti/LaTeX/Bibliography}

\bsp

\label{lastpage}

\end{document}